\documentclass[twocolumn]{aastex701}

\begin{document}

\title{Two Potential Exoplanets around A-type Stars Selected from 18 Planetary Candidates}

\author[0009-0001-3899-1669]{Yi-fan Luo}
\affiliation{Yunnan Observatories, Chinese Academy of Sciences (CAS), 650216 Kunming, China}
\affiliation{University of Chinese Academy of Sciences, No.1 Yanqihu East Rd, Huairou District, Beijing, China 101408}
\email{luoyifan@ynao.ac.cn}

\author[0000-0001-9346-9876]{Wen-Ping Liao}
\affiliation{Yunnan Observatories, Chinese Academy of Sciences (CAS), 650216 Kunming, China}
\affiliation{University of Chinese Academy of Sciences, No.1 Yanqihu East Rd, Huairou District, Beijing, China 101408}
\email[show]{liaowp@ynao.ac.cn}
\correspondingauthor{Wen-Ping Liao}

\author[0000-0002-5995-0794]{Sheng-Bang Qian}
\affiliation{School of Physics and Astronomy, Yunnan University, Kunming 650091, China}
\email{qiansb@ynu.edu.cn}

\author[0000-0002-8276-5634]{Wen-Xu Lin}
\affiliation{School of Physics and Astronomy, Yunnan University, Kunming 650091, China}
\email{linwenxu@ynu.edu.cn}

\author[0000-0002-0796-7009]{Li-Ying Zhu}
\affiliation{Yunnan Observatories, Chinese Academy of Sciences (CAS), 650216 Kunming, China}
\affiliation{University of Chinese Academy of Sciences, No.1 Yanqihu East Rd, Huairou District, Beijing, China 101408}
\email{zhuly@ynao.ac.cn}

\begin{abstract}
We screen and analyze exoplanet candidates around A-type stars
(defined as $T_{\mathrm{eff}}$ between 7500 and 10000~K) observed by
the Transiting Exoplanet Survey Satellite (TESS) to evaluate their likelihood
of being genuine exoplanets. 
Our analysis involves transit signal searches, light-curve 
detrending, estimation of nearby-source contamination, and calculation of false positive probabilities (FPPs). 
Among the 18 candidates analyzed, 
four exhibit relatively low FPP values ($<$15\%).
Two candidates are excluded
from further analysis due to the lack of stellar parameter data. 
Six candidates show no clearly detectable transit signals, likely due to shallow or weak features, 
while six candidates exhibit relatively high FPP values, leaving their authenticity uncertain.
Among the four low-FPP targets, two—TIC~48031665 and TIC~259230140—stand out as the most promising.
TIC~48031665 shows a very shallow transit 
signal but has a very low FPP and minimal nearby starlight contamination.
TIC~259230140 displays a clear U-shaped transit light curve typical of
planetary transits, along with slightly higher yet still low FPP and contamination
levels. These two objects are therefore considered the most promising candidates
identified in this study.
\end{abstract}

\section{Introduction}

Exoplanet research is one of the fastest-developing frontier fields in 
contemporary astrophysics. So far, more than 6000 exoplanets have 
been confirmed; the vast majority of which orbit late-type, low-mass 
stars \citep{exoplanet_web_2025}.  In contrast, planetary systems 
around early-type stars, especially A-type stars, are still very scarce.

\begin{figure*}[ht!]
    \centering
    \includegraphics[width=0.95\textwidth]{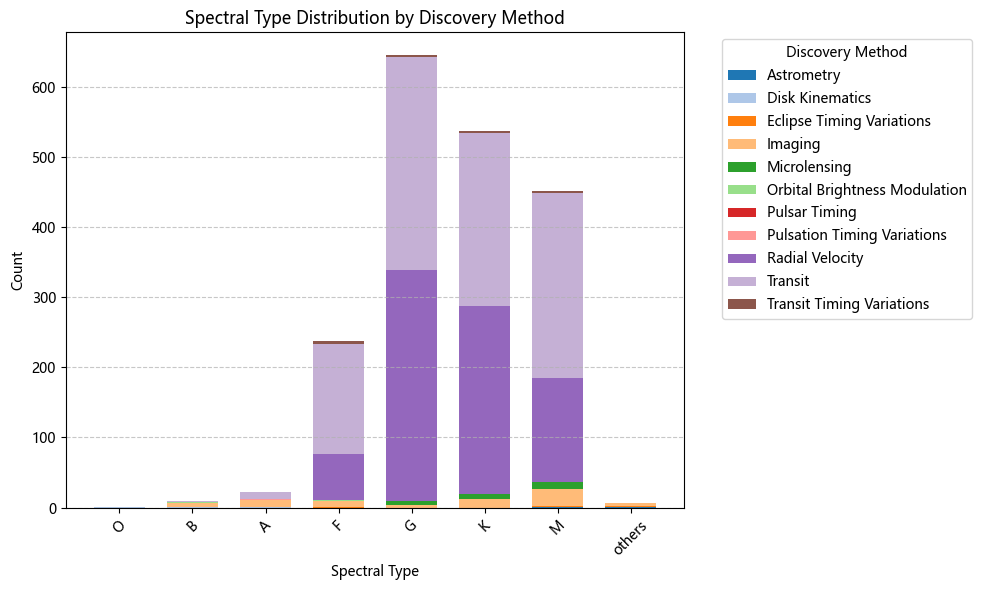}
    \caption{Statistical distribution of exoplanets among host stars of different spectral types and discovery methods.  
    The data come from the NASA Exoplanet Archive.  
    It can be seen that most exoplanets are discovered around G–M type stars, while the number of A-type star planets is significantly smaller.}
    \label{fig:spectral_type_distribution}
\end{figure*}

The effective temperature of A-type stars ranges from 7500 to 10000 K, 
and their masses are usually 1.4-2.1 times that of the Sun\citep{Pecaut_2013, Gray_2021}.  
They are characterized by high luminosity, rapid rotation, and relatively 
short lifetimes \citep{Royer_2007}.  
These characteristics make them challenging targets in exoplanet studies.  
On the one hand, the weak magnetic fields, broadened absorption lines, and 
possible pulsations of A-type stars make their radial velocity (RV) 
measurements extremely difficult \citep{Becker_2015};  
on the other hand, their large radii dilute the transit depth, making 
the transit signals more difficult to detect \citep{Johnson_2011}.  
As a result, only about 20 exoplanets orbiting A-type stars have been 
discovered so far \citep{Marois_2008, Hartman_2015, Borgniet_2019}.  
To show the current distribution of exoplanets among stars of different 
spectral types, we performed statistics on the host star types and discovery 
methods based on the NASA Exoplanet Archive 
\footnote{\url{https://exoplanetarchive.ipac.caltech.edu/}} 
(see Figure~\ref{fig:spectral_type_distribution}).  
The results show that most exoplanets are found around G–M type stars, while 
A-type star samples are clearly scarce, highlighting the need for systematic research on them.

As shown in Figure~\ref{fig:spectral_type_distribution}, the detectability of exoplanets 
strongly depends on stellar spectral type and observational method. 
For O and B-type stars, their extremely high luminosities and large stellar radii 
significantly dilute transit depths, making transit signals particularly shallow and 
difficult to detect. To date, only one confirmed transiting planet has been detected 
around a B-type star, KELT-9\,b \citep{Gaudi_2017}, highlighting the observational challenges 
in this regime. In contrast, although F-type stars are generally more favorable for transit detections, 
their physical properties are closer to those of Sun-like (G-type) stars. 
Therefore, studying planets around A-type stars provides greater leverage for exploring 
planet formation and evolution under distinctly different stellar 
conditions. A-type stars thus represent a critical intermediate regime, where both transit 
and direct imaging detections are feasible, offering a unique opportunity to study planetary 
systems around hot, massive stars while maintaining a reasonable detection efficiency.

Despite these difficulties, the study of A-type star planets still holds unique scientific value.  
Their high luminosity is favorable for high-resolution atmospheric observations \citep{Seager_2010},  
and their massive protoplanetary disks may result in formation and migration mechanisms different from those of Sun-like stars \citep{Ida_2005, Kennedy_2008}.  
Typical systems such as HR~8799 \citep{marois_2010} demonstrate the potential of direct imaging.  
In recent years, large-scale survey projects such as KELT \citep{Pepper_2003, Pepper_2012},  
MASCARA \citep{Talens_2017}, and Transiting Exoplanet Survey Satellite (TESS) \citep{TESS_article_2014}  
have greatly expanded the possibility of conducting transit searches around bright, early-type stars.  
However, because the TESS pixel size is relatively large (about $21^{\prime\prime}$), 
a transit signal may originate from an eclipsing binary or a background star 
hosting an eclipsing system, rather than from a true planetary transit.
Therefore, it is necessary to introduce robust statistical methods to evaluate the reliability of A-type star candidates.

In this study, based on TESS observational data, we conducted a survey and analysis 
of a group of candidate planets around A-type stars  
(defined here as stars with effective temperatures between 7500 and 10000 K).  
By integrating light-curve modeling, prior parameters of the host stars, 
and tools such as \texttt{TESS-Cont} \citep{tess-cont-cit}
and \texttt{TRICERATOPS} \citep{triceratops_ARTICLE, triceratops_package},  
we classified and evaluated the reliability of these candidates and 
identified two promising targets for follow-up observations.

Unlike the large-sample statistical study by \citet{triceratops_ARTICLE},  
this work focuses on high-temperature host star samples with $T_{\mathrm{eff}} = 7500$–$10000$~K.  
By combining the contamination analysis from \texttt{TESS-Cont} with the 
false positive probability (FPP) and the nearby false positive probability (NFPP)  
estimation from \texttt{TRICERATOPS},  
we carried out a more detailed individualized vetting of these targets.  
The structure of this paper is as follows:
Section 2 introduces the data sources and sample selection criteria;
Section 3 describes the data processing methods and presents the calculations and results of the FPP, NFPP, and nearby starlight contamination;
Section 4 classifies and discusses the analyzed samples;
and Section 5 provides the conclusions and discussions.

\section{Data and Target Samples}

\subsection{Data Sources}

The observational data used in this study were obtained from TESS \citep{TESS_article_2014}.  
Since its launch in 2018, TESS has continuously conducted high-precision photometric 
monitoring of bright stars across the entire sky,  
providing multiple observing cadences (e.g., 120\,s, 200\,s, 600\,s, and 1800\,s).  
We used the \texttt{lightkurve} package \citep{lightkurve_package} to download the light 
curves and target pixel files (TPFs) for each target.  
For each source, we selected light curves with a uniform observing cadence and 
a clearly identifiable transit signal after detrending, which were suitable for 
subsequent light-curve modeling and parameter estimation. 
Light curves that led to unstable detrending or failed to converge during the MAP optimization 
were excluded from further analysis.

In the process of light-curve analysis, the basic physical parameters of the host stars (e.g., mass and radius) were also required.  
These parameters were obtained from the ExoFOP-TESS database\footnote{\url{https://exofop.ipac.caltech.edu/tess/}},  
which integrates information from the TESS Input Catalog (TIC) and Gaia,  
providing the necessary priors for transit modeling and false positive probability calculations.  

\subsection{Sample Selection Criteria}

We applied the following selection criteria to the TESS Objects of Interest (TOIs) listed in the NASA Exoplanet Archive:

\textbf{(1) Effective temperature between 7500 and 10000 K:}  
This range corresponds to typical A-type stars.  
Stars with higher temperatures are usually brighter and exhibit shallower transit signals,  
and their spectra are strongly affected by rotational broadening.  
Together with their larger stellar masses, these factors make precise radial-velocity measurements more challenging \citep{Royer_2007, Becker_2015}.  
Stars with lower temperatures fall outside the scope of this study.  

\textbf{(2) Orbital period longer than 1 day:}  
Although this criterion excludes ultra-short-period planets (USPs),  
their occurrence rate around G-type stars is only about 0.5\% \citep{Sanchis-Ojeda_2014}. 
In contrast, a significant fraction of eclipsing binary systems are known to have 
orbital periods shorter than 1 day \citep{Slawson_2011}.  
Therefore, we infer that companions around A-type stars with periods 
shorter than 1 day are more likely to be stellar rather than planetary in nature. 
This condition effectively helps eliminate potential eclipsing binaries while retaining 
the majority of true planetary candidates. 

\begin{figure}[ht!]
    \centering
    \includegraphics[width=\linewidth]{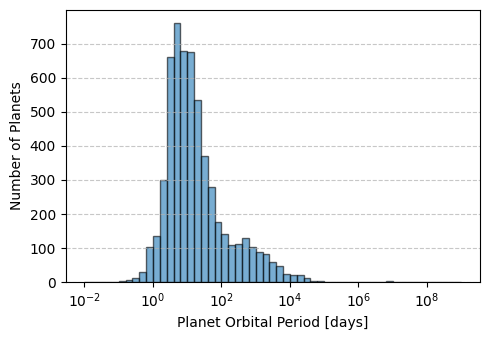}
    \caption{Orbital period distribution of confirmed planets from the 
    NASA Exoplanet Archive (as of September~29,~2025). 
    It can be seen that most confirmed planets have orbital periods longer than 1~day. 
    }
    \label{fig:period_distribution}
\end{figure}

Our statistical analysis based on confirmed planets from the  
NASA Exoplanet Archive (Figure~\ref{fig:period_distribution})  
shows that most planetary orbits are longer than 1 day. 

\textbf{(3) Planetary radius smaller than 7\,R$_\oplus$:}  
Brown dwarfs typically have radii comparable to or slightly smaller than Jupiter's,  
generally in the range of 9--13\,R$_\oplus$ \citep[e.g.,][]{Burrows_1997, Baraffe_2002}.  
We therefore adopted a stricter upper limit of 7\,R$_\oplus$ to effectively exclude non-planetary scenarios  
and ensure the purity of the candidate sample.  

\textbf{(4) TFOPWG disposition marked as PC or APC:}  
The TESS Follow-up Program Working Group (TFOPWG) provides disposition labels including  
APC (Ambiguous Planetary Candidate), FA (False Alarm), FP (False Positive),  
KP (Known Planet), PC (Planetary Candidate), and CP (Confirmed Planet).  
For the purpose of this study, we prioritized targets labeled as PC  
and additionally included two sources labeled as APC to further examine their reliability.  

After applying these criteria, we obtained a total of 18 exoplanet candidates orbiting A-type stars,  
which are analyzed in detail in the following sections.
A summary of their stellar and planetary parameters is provided in Table~\ref{tab:pc_summary} in the Appendix.

\section{Data Processing and Results}

We processed the acquired data as follows. 
First, we applied the Box Least Squares (BLS) algorithm to search for the orbital period of 
transiting signals 
and estimated the system parameters using the Maximum A Posteriori (MAP) optimization method. 
The light curves were then detrended and phase-folded. 
These steps follow the procedures outlined in the \textsf{exoplanet} 
tutorials\footnote{\url{https://gallery.exoplanet.codes/tutorials/tess/}}. 
Subsequently, we input the relevant parameters and the phase-folded light curves 
into the \texttt{TRICERATOPS} package 
to calculate the FPP and the NFPP. 
In addition, the \texttt{TESS-Cont} package was used to estimate flux contamination 
from nearby stars. 
The detailed procedures and corresponding results are described below.

\begin{deluxetable*}{lcccccc}
\tablecaption{Comparison between MAP and MCMC results (Example Targets) \label{tab:map_mcmc_params}}
\tablehead{
\colhead{TIC ID} & \colhead{Method} & \colhead{Period (days)} & \colhead{Depth} &
\colhead{$t_0$} & \colhead{FPP (\%)} & \colhead{NFPP (\%)}
}
\startdata
259230140 & MAP  & 14.31698 & 0.00162 & -375.094 & $6.96 \pm 1.02$ & $0.14 \pm 0.02$ \\
          & MCMC & $14.31708 \pm 0.00016$ & $0.00159 \pm 0.00012$ & $-375.099 \pm 0.006$ & $8.30 \pm 1.04$ & $0.19 \pm 0.03$ \\
\hline
48031665  & MAP  & 13.11869 & 0.00036 & -564.636 & $2.28 \pm 0.03$ & $0.18 \pm 0.00$ \\ 
          & MCMC & $13.11897 \pm 0.00029$ & $0.00037 \pm 0.00003$ & $-564.671 \pm 0.014$ & $2.38 \pm 0.03$ & $0.19 \pm 0.00$ \\ 
\hline 
22069559  & MAP  & 2.03422 & 0.00025 & -13.1722 & $15.45 \pm 0.03$ & $2.73 \pm 0.00$ \\
          & MCMC & $2.03409 \pm 0.00085$ & $0.00025 \pm 0.00003$ & $-13.172 \pm 0.010$ & $15.43 \pm 0.04$ & $2.71 \pm 0.01$ \\
\enddata
\tablecomments{
All MCMC chains achieved satisfactory convergence ($\hat{R} < 1.05$).
For these three representative targets, the MAP-derived parameters are broadly
consistent with the MCMC posterior medians, showing similar overall trends.
This consistency supports the reliability of the MAP solutions for
phase-folding and subsequent analyses, while offering a more efficient
alternative to full MCMC sampling. Note that the MAP method provides only
the maximum a posteriori estimates and does not yield parameter uncertainties.
}
\end{deluxetable*}

\subsection{Data Processing}

The BLS algorithm \citep{Kovacs2002} 
fits a series of trial box-shaped models to the light curve over a range of test periods 
and trial transit times (epochs)
and identifies the period and epoch that minimizes the residual sum of squares. 
This method is particularly effective for detecting short-duration, 
low-amplitude, and nearly rectangular transit-like signals.

Since the true transit signal of an exoplanet does not always correspond 
to the strongest BLS power peak, 
we referred to the candidate periods listed in the NASA Exoplanet Archive 
and searched for local maxima of the BLS power near those values 
to obtain preliminary estimates of the orbital period and mid-transit time. 
These parameters, together with stellar mass and radius information from the ExoFOP database, 
served as inputs for subsequent modeling.

During the modeling stage, the light curve was represented 
as the sum of a planetary transit signal and a background trend 
described by a Gaussian Process (GP) model. 
The GP simultaneously captures both systematic variations and intrinsic stellar variability, 
allowing for detrending without compromising the transit signal. 
We then employed the MAP optimization method to determine the most probable 
system parameters under the given priors. 
In this process, several candidates did not exhibit clear transit signals 
and were therefore excluded from further analysis; 
these cases are discussed in the next section.

Although in principle a full Markov Chain Monte Carlo (MCMC) sampling 
should be used to obtain the posterior distributions of parameters, 
the computational cost of MCMC is high. 
To evaluate the consistency between the two approaches, 
we compared MAP and MCMC results for three representative candidates 
(see Table~\ref{tab:map_mcmc_params}). 
The results show that the differences between the two methods 
in the subsequent FPP and NFPP calculations are acceptable. 
Therefore, this study primarily adopts MAP-optimized parameters 
to balance computational efficiency and reliability.

\subsection{FPP and NFPP Estimation}

\texttt{TRICERATOPS} models the likelihood of different astrophysical scenarios based 
on the observed transit features of the target and information on nearby sources 
from stellar catalogs, including scenarios such as True Planet (TP), Eclipsing 
Binary (EB), Eclipsing Binary with $2\times Period$ (EBx2P), Secondary Eclipsing 
Binary (SEB), and Background Eclipsing Binary (BEB). The outputs include 
the posterior probabilities for each scenario, as well as the overall FPP and NFPP.
A summary of all the astrophysical scenarios considered in the \texttt{TRICERATOPS} 
framework is provided in Table~\ref{tab:triceratops_scenarios}.

\begin{deluxetable}{ll}
\tablecaption{Astrophysical Scenarios Tested by \texttt{TRICERATOPS} \label{tab:triceratops_scenarios}}
\tablehead{
\colhead{Abbrev.} & \colhead{Scenario Type}
}
\startdata
TP & True Planet \\
EB & Eclipsing Binary \\
EBx2P & Eclipsing Binary with $2\times Period$ \\
PTP & Primary Transiting Planet \\
PEB & Primary Eclipsing Binary \\
PEBx2P & Primary Eclipsing Binary with $2\times Period$ \\
STP & Secondary Transiting Planet \\
SEB & Secondary Eclipsing Binary \\
SEBx2P & Secondary Eclipsing Binary with $2\times Period$ \\
DTP & Diluted Transiting Planet \\
DEB & Diluted Eclipsing Binary \\
DEBx2P & Diluted Eclipsing Binary with $2\times Period$ \\
BTP & Background Transiting Planet \\
BEB & Background Eclipsing Binary \\
BEBx2P & Background Eclipsing Binary with $2\times Period$ \\
NTP & Nearby Transiting Planet \\
NEB & Nearby Eclipsing Binary \\
NEBx2P & Nearby Eclipsing Binary with $2\times Period$ \\
\enddata
\tablecomments{List of the 18 astrophysical scenarios considered in \texttt{TRICERATOPS} \citep{triceratops_ARTICLE}.}
\end{deluxetable}

According to the criteria of \citet{triceratops_ARTICLE}, a candidate can be 
statistically validated as a planet (validated planet) if $\mathrm{FPP} < 1.5\%$ and $\mathrm{NFPP} < 0.1\%$. 
It should be noted that the results depend on 
the quality of the light curve and the completeness of the stellar catalog 
for nearby sources, and the modeling assumptions may introduce uncertainties. 
Therefore, for candidates with borderline values (FPP neither very low nor 
very high), the final assessment still requires combining transit morphology 
with other observational evidence. Considering the stochastic nature of the 
calculations, each target was independently processed at least 20 times, and 
the mean of these runs was adopted as the final output. 

Some targets could not have their FPP and NFPP computed for the following 
reasons.
First, TIC~422159302 and TIC~53709089 lack stellar mass information, 
making MAP fitting and detrending impossible, 
and thus FPP/NFPP could not be obtained. Second, some 
candidates have very shallow transits, causing MAP optimization to fail to 
converge. 
Third, for certain targets, 
the observed stellar variability period may 
have been misidentified as a transit signal (consistent with the candidate 
period listed in the NASA Exoplanet Archive), resulting in no detected transit 
and preventing subsequent MAP fitting and FPP/NFPP calculation. 

Among the 10 targets for which FPP could be computed, 2 have FPP values in 
the 40-50\% range, 3 in 20-40\%, 2 in 10-20\%, and the remaining 3 in 2-10\%. 
All computed results are listed in Table~\ref{tab:pollution}, and detailed 
analysis and discussion are presented in the next subsection. 

\subsection{Contamination Analysis}

\begin{deluxetable*}{lccccc}
\tablecaption{Results of contamination and false positive probability analyses \label{tab:pollution}}
\tablehead{\colhead{TIC ID} & \colhead{Contam. (\%)} & \colhead{Star 1 Contam. (\%)} &
\colhead{FPP (\%)} & \colhead{NFPP (\%)} & \colhead{Notes}}
\startdata
422159302 & $28.8 \pm 1.9$  & $61.5 \pm 6.4$  & \nodata & \nodata & Missing stellar mass \\
400972123 & $10.2 \pm 1.0$  & $42.6 \pm 3.1$  & $20.44 \pm 0.03$ & $5.92 \pm 0.01$ & \nodata \\
378613125 & $9.8 \pm 0.4$   & $32.0 \pm 1.9$  & $29.34 \pm 0.06$ & $4.67 \pm 0.01$ & \nodata \\
367439938 & $13.6 \pm 0.5$  & $70.2 \pm 3.7$  & $11.96 \pm 0.05$ & $3.21 \pm 0.01$ & \nodata \\
350575997 & $10.0 \pm 1.0$  & $41.0 \pm 2.0$  & $40.80 \pm 0.16$ & $5.25 \pm 0.02$ & \nodata \\
340458804 & $3.0 \pm 1.6$   & $50.0 \pm 15.7$ & \nodata & \nodata & Too shallow for fitting \\
259353953 & $8.1 \pm 0.9$   & $33.7 \pm 3.1$  & $31.94 \pm 0.10$ & $3.47 \pm 0.01$ & \nodata \\
259230140 & $6.8 \pm 0.9$   & $16.0 \pm 1.8$  & $6.96 \pm 1.02$   & $0.14 \pm 0.02$ & \nodata \\
231721005 & $6.8 \pm 0.5$   & $86.3 \pm 4.1$  & \nodata & \nodata & No clear signal \\
207277638 & $1.4 \pm 0.2$   & $29.0 \pm 2.3$  & \nodata & \nodata & APC, variability dominated \\
180575165 & $7.0 \pm 1.0$   & $23.4 \pm 1.1$  & $48.55 \pm 0.04$ & $4.65 \pm 0.00$ & \nodata \\
166086403 & $0.017 \pm 0.005$ & $45.4 \pm 7.2$ & \nodata & \nodata & Too shallow for fitting \\
126737992 & $0.28 \pm 0.04$ & $38.2 \pm 5.6$  & \nodata & \nodata & APC, MAP failed \\
60984804  & $0.03 \pm 0.00$ & $22.0 \pm 2.4$  & $2.98 \pm 0.14$  & $0.00 \pm 0.00$  & V-like transit shape \\
53709089  & $45.6 \pm 0.0$  & $93.0 \pm 0.0$  & \nodata & \nodata & Too contaminated \\
48031665  & $0.49 \pm 0.07$ & $21.9 \pm 2.0$  & $2.28 \pm 0.03$  & $0.18 \pm 0.00$ & \nodata \\
32498058  & $6.3 \pm 1.2$   & $96.0 \pm 0.8$  & \nodata & \nodata & No clear signal \\
22069559  & $5.0 \pm 0.3$   & $16.6 \pm 0.5$  & $15.45 \pm 0.03$ & $2.73 \pm 0.00$  & \nodata \\
\enddata
\tablecomments{
Contam. (\%) and Star~1~Contam. (\%) are derived from \texttt{TESS-Cont},
representing the total contamination ratio and the fractional flux contribution
from the brightest contaminating source within the TESS aperture, respectively.
FPP (false positive probability) and NFPP (nearby false positive probability) 
are the mean values of 20 independent \texttt{TRICERATOPS} runs,
with $1\sigma$ standard deviations quantifying the run-to-run variability.
}
\end{deluxetable*}

Since the spatial resolution of TESS is not very high (the pixel scale is approximately $21^{\prime\prime}$),
observations of target sources are easily affected by nearby stars.
This may lead to signals from adjacent foreground or background stars being mixed into the target's aperture pixels,
which could cause the transit signal of a nearby star to be mistakenly attributed to the target star.
Although \texttt{TRICERATOPS} already considers such scenarios (e.g., Background Transiting Planet, BTP; 
BEB), and provides corresponding probabilities,
the stellar contamination estimates provided by \texttt{TESS-Cont} help to more intuitively assess the likelihood of these contamination scenarios.

\texttt{TESS-Cont} is a Python-based tool that operates by reading an \texttt{.ini} file containing the target information.
It automatically calculates the fractional flux contributions from the target itself and nearby sources
within the corresponding aperture pixels. 
It then reports the flux contribution ratios of the most contaminating neighboring stars.
Since the observing conditions vary between different sectors,
we performed \texttt{TESS-Cont} analyses on the TPFs of all targets and carried out statistical assessments.
Table~\ref{tab:pollution} lists the mean contamination fractions and their uncertainties for all 18 sources.

The results show that most targets have contamination fractions below 15\%, with only a few exceeding 20\%,
indicating that their transit signals mainly originate from the host stars themselves.
Across different sectors, the absolute difference in contamination fraction is typically less than two percentage points,
demonstrating that the results are relatively robust.
Only a few targets (such as TIC~340458804) exhibit large relative differences between sectors,
which can be attributed to the aperture boundary effect caused by nearby bright stars—
that is, part of the flux from the bright star is inconsistently included or excluded
in the aperture mask selection at different observation times,
leading to significant sector-to-sector variation in contamination fraction.
A few sources, such as TIC 53709089 and TIC 422159302,
show notably high contamination levels, suggesting that the authenticity of their transit signals should be evaluated with caution.

\section{Case Studies}

Based on the results presented in the previous sections, the 18 targets analyzed in this study 
can be classified into the following categories:  
(1) three targets exhibiting low FPP values and shallow transit signals but with differing levels of reliability;  
(2) one promising candidate that shows a relatively deep and well-defined \text{U-shaped} transit;  
(3) six targets in which no clear transit signals were detected near the reported periods;  
(4) six targets showing relatively high FPP values ($>15\%$) and shallow transit depths, 
for which the observed signals may originate either from planetary transits or eclipsing binaries.  

In addition, TIC~422159302 and TIC~53709089 were 
excluded from the FPP/NFPP analysis 
due to the lack of reliable host-star mass estimate.  
By jointly examining the light-curve morphology and the FPP results, 
we aim to more clearly assess the credibility of each candidate 
and identify the potential sources of uncertainty in their classification.

\subsection{Low-FPP Shallow-Transit Targets}

\begin{figure*}[ht!]
\gridline{\fig{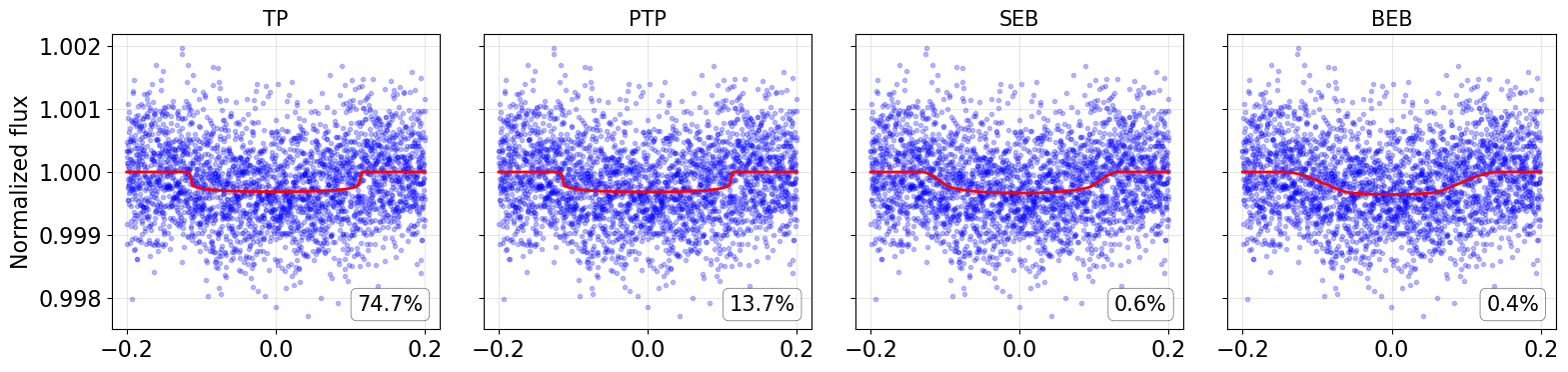}{\textwidth}{(a) TIC 48031665}}
\gridline{\fig{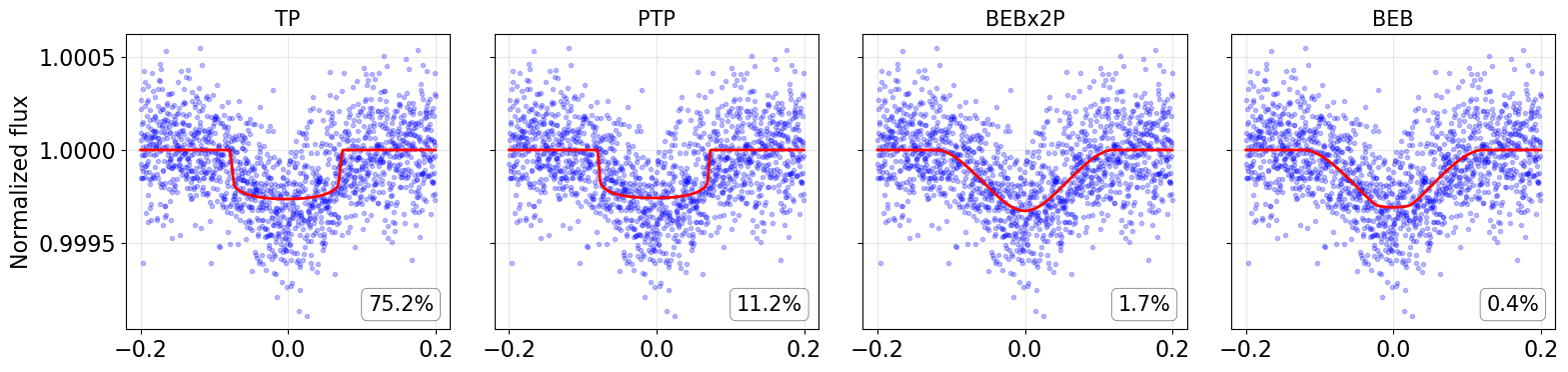}{\textwidth}{(b) TIC 60984804}}
\gridline{\fig{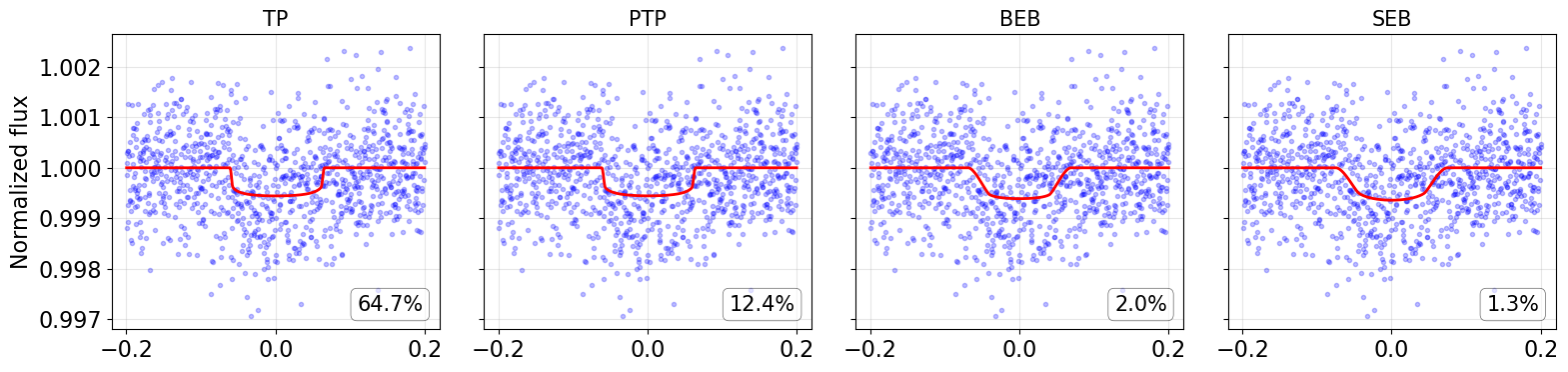}{\textwidth}{(c) TIC 367439938}}
\caption{Phase-folded transit light curves for three targets with relatively 
low false positive probabilities (FPP $<15\%$). 
Each panel shows the most probable planetary and eclipsing binary scenarios, 
with lower-right labels indicating their probabilities. 
TIC~48031665 exhibits a shallow transit-like signal; given its very low FPP (2.28\%), 
the likelihood of a binary origin is minimal, making it a highly promising candidate. 
TIC~60984804 also shows a low FPP (2.98\%) but presents a morphology more consistent 
with an eclipsing binary, warranting caution in its interpretation. 
TIC~367439938 shows no clear distinction between planetary (U-type) 
and binary (V-type) morphologies, leaving its nature ambiguous.}
\label{fig:low_fpp_cases}
\end{figure*}

Within the sample, three sources — TIC~48031665, TIC~60984804, and TIC~367439938 — exhibit low FPP ($<15\%$). 
Their detailed model fits under both planetary and eclipsing-binary scenarios 
are presented in Figure~\ref{fig:low_fpp_cases}.

For TIC~48031665, the transit signal was derived from multiple sectors of 120\,s 
cadence data.  
After detrending and phase-folding, the transit depth is relatively shallow 
but still discernible (Figure~\ref{fig:low_fpp_cases}a).  
\texttt{TRICERATOPS} produced FPP = $(2.28 \pm 0.03)\%$ and 
NFPP = $(0.18 \pm 0.00)\%$.  
Among the scenario probabilities, the TP case dominates 
($\sim75\%$), followed by primary transiting planet (PTP, $\sim14\%$) and 
Diluted Transiting Planet (DTP, $\sim7\%$) scenarios.  
For eclipsing binary scenarios, SEB is the most probable, with a probability of only $\sim0.6\%$.  
Meanwhile, \texttt{TESS-Cont} yielded a contamination ratio of 
$(0.49 \pm 0.07)\%$, suggesting a relatively clean stellar field 
with negligible flux contribution from nearby stars.  
Taken together, the transit signal for this target is highly credible 
and warrants follow-up observations.  

By contrast, TIC~60984804 also shows a low false positive probability 
(FPP = $(2.98 \pm 0.14)\%$, NFPP $< 10^{-4}$),  
yet its phase-folded light curve displays an ambiguous, nearly V-shaped morphology  
(Figure~\ref{fig:low_fpp_cases}b), which is more characteristic of eclipsing binaries.  
In the scenario probabilities, the TP remains dominant ($\sim75\%$),  
while the probabilities for eclipsing binary scenarios are higher than in the previous case but still low,
such as Background Eclipsing Binary with $2\times Period$ 
(BEBx2P, $\sim1.7\%$) and BEB ($\sim0.4\%$).  
However, as Figure~\ref{fig:low_fpp_cases}b shows, its folded light curve 
is visually more consistent with the eclipsing binary scenarios than with the planetary ones.
Thus, despite its low numerical FPP, this target should be interpreted with caution,  
as it may not represent a bona fide planetary transit.  

The third case is TIC~367439938 (Figure~\ref{fig:low_fpp_cases}c),  
which shows a contamination fraction of $(13.6\pm0.5)\%$,  
FPP = $(11.96 \pm 0.05)\%$, and NFPP = $(3.21 \pm 0.01)\%$.  
Unlike the previous two cases that were based on 120\,s cadence data,
this target was observed in 600\,s cadence mode,
resulting in fewer data points and lower time resolution.
Although these values remain below 15\%, the observed light curve is difficult to classify,
as it shows no clear distinction between U-shaped and V-shaped morphologies.
Moreover, the fitted curves for planetary (TP, PTP) and eclipsing binary (BEB, SEB) scenarios
are comparably consistent with the data, making it challenging to determine the true nature of the signal.
Therefore, this target remains ambiguous and warrants additional follow-up observations for confirmation.

In summary, TIC~48031665, TIC~60984804, and TIC~367439938 represent three targets 
that exhibit similarly shallow transit signals and low FPP values, yet with 
markedly different levels of reliability.  
Among them, TIC~48031665 appears to be a strong planetary candidate 
among the three; 
however, all three targets still require additional follow-up observations 
to confirm 
whether their signals originate from genuine exoplanetary transits or eclipsing binaries.  
These cases highlight the limitation of using FPP as a 
sole diagnostic criterion for planet validation---particularly when 
FPP exceeds 1.5\%---and underscore the necessity of combining transit morphology 
with complementary observational evidence for a more robust assessment.

\subsection{Low-FPP U-shaped transit Target}

\begin{figure}[htbp]
    \centering
    \includegraphics[width=1.0\linewidth]{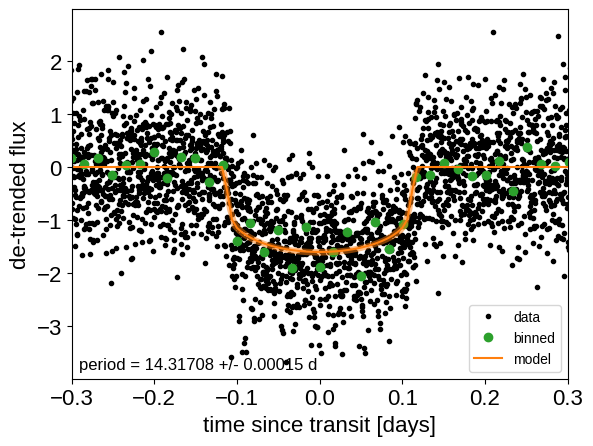}
    \caption{Phase-folded transit light curve of TIC~259230140 (black points).  
    The green points indicate binned averages, and the orange curve represents 
    the best-fit model.  
    The clear U-shaped transit profile, 
    together with its low NFPP and moderate contamination, 
    indicates that this target is a high-quality planetary candidate.
    The vertical axis is in units of $\times10^{-3}$.}
    \label{fig:tic259230140}
\end{figure}

In the sample of 18 targets studied here, we note that TIC~259230140 
exhibits a pronounced transit signal, displaying a clear U-shaped light 
curve typical of planetary transits (Fig.~\ref{fig:tic259230140}). 
\texttt{TRICERATOPS} yields FPP = $(6.96 \pm 1.02)\%$ and NFPP = $(0.14 \pm 0.02)\%$, 
while \texttt{TESS-Cont} reports a contamination fraction of $6.8 \pm 0.9\%$.  
Although the FPP has not met the commonly used confirmation threshold of 1.5\%, 
its value below 10\% combined with the clear U-shaped light curve 
suggests a relatively high likelihood of being an exoplanet.  
Considering the relatively low contamination and extremely small NFPP, 
this target shows significant potential as a candidate.  
According to the NASA Exoplanet Archive, the candidate has a radius 
of approximately $6.38 \pm 0.39\,\text{R}_\oplus$
and an orbital period of about 14.317\,days.  
We strongly recommend follow-up work, including radial velocity 
measurements and complementary photometric observations, 
to further confirm its exoplanetary nature.

\subsection{Targets Without Detected Transit Signals}

Although the NASA Exoplanet Archive provides the reported transit depths, orbital periods, 
and other parameters for each candidate, we found that several targets in our sample 
did not exhibit any convincing transit signal at the listed orbital periods. 
These sources can be broadly divided into two categories.

\begin{figure}[htbp]
    \centering
    \includegraphics[width=1.0\linewidth]{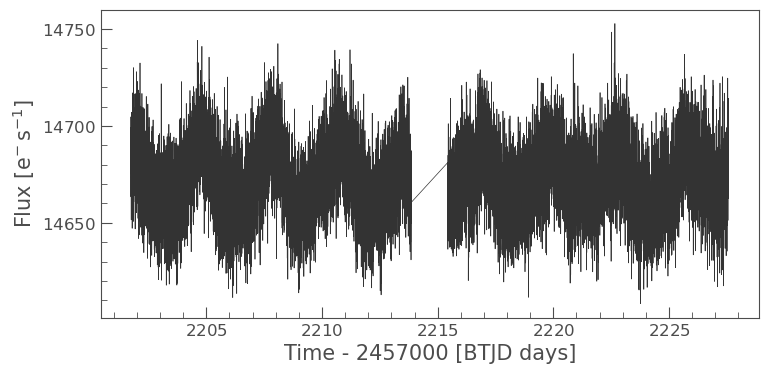}
    \caption{Segment of the light curve for TIC~207277638 showing quasi-periodic 
    variations likely dominated by intrinsic stellar variability. 
    No convincing transit-like features are detected in the data.}
    \label{fig:apc_lc}
\end{figure}

\begin{figure}[htbp]
  \centering
  \includegraphics[width=\linewidth]{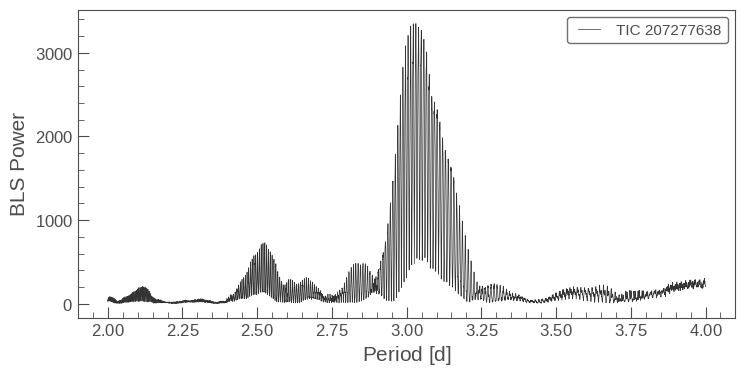}
  \includegraphics[width=\linewidth]{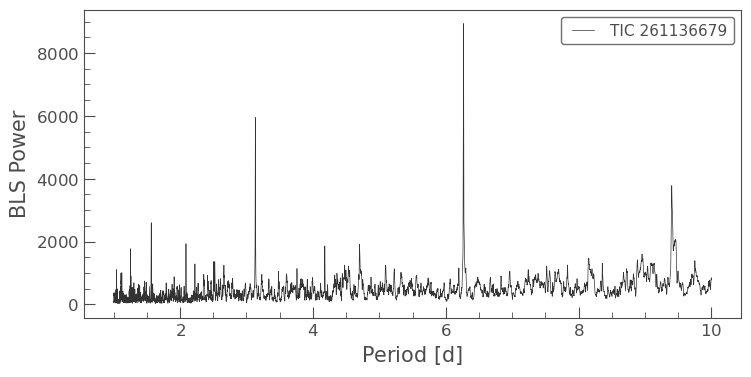}
  \caption{Comparison of BLS periodograms. 
  The upper panel shows TIC~207277638, listed in the NASA Exoplanet Archive with a reported 
  candidate period of 3.121\,days, exhibiting a broad peak near that value, 
  indicating variability inconsistent with a planetary transit.
  The lower panel shows TIC~261136679, a confirmed planet \citep{Huang_2018}, 
  which is reanalyzed here for comparison.
  It exhibits a sharp and isolated peak at 6.27\,days, 
  characteristic of a genuine transiting planet signal. 
  Both periodograms were produced in this work using the BLS algorithm.}
  \label{fig:apc_bls}
\end{figure}

The first category (TIC~207277638, TIC~126737992)
comprises targets whose light curves exhibit clear periodic modulations
that coincide with the reported orbital periods listed in the NASA Exoplanet Archive.
These light curves show quasi-periodic variations near the reported periods.
We suspect that such signals are more likely caused by intrinsic stellar variability
that was misidentified as transit-like features during the initial automated vetting process.
To further test for possible transit events, we performed a BLS search.
Ultimately, no credible transit signatures were found at the expected periods
for any of these three targets.
As an example, Figure~\ref{fig:apc_lc} shows a segment of the light curve for TIC~207277638.
Its reported period is about 3.12\,days,
but the corresponding BLS spectrum exhibits a broad, non-sharp power peak
(Figure~\ref{fig:apc_bls}, top), in contrast to the well-defined peak typically seen
in genuine planetary transits (Figure~\ref{fig:apc_bls}, bottom).

The second category comprises TIC~166086403, TIC~340458804, TIC~231721005, 
and TIC~32498058. Among them, the signals of TIC~166086403 and TIC~340458804 
are too weak to converge during the MAP optimization process, 
preventing reliable detrending or period estimation. 
Meanwhile, the light curves of TIC~231721005 and TIC~32498058 remain largely flat, 
showing no clear transit-like events near the reported periods. 
As a result, none of these targets can be confidently identified as hosting transiting planets.

Although these sources do not provide robust evidence of transits, 
they still hold scientific value. On the one hand, such ``non-detections'' 
serve as exclusionary references for future observations, 
helping to avoid unnecessary allocation of limited observing resources to low-confidence targets. 
On the other hand, by comparing these cases with successfully identified candidates, 
we can better understand the limitations of current transit search algorithms 
and gain insights that may guide future improvements in signal filtering and modeling strategies.

\subsection{Marginal or Uncertain Candidates}

Except for the individual cases discussed above, the transit signals of the remaining 
candidates are relatively weak and poorly defined, making it difficult to determine 
whether their light curves correspond to the typical U-shaped profiles of planetary transits, 
the V-shaped signatures of eclipsing binaries, or other transit-like variations.  
The FPP values provided by \texttt{TRICERATOPS} are all greater than 15\%.  
These targets include TIC~400972123 (FPP $\approx 20\%$), TIC~378613125 (FPP $\approx 29\%$),  
TIC~350575997 (FPP $\approx 41\%$), TIC~259353953 (FPP $\approx 32\%$),  
and TIC~180575165 (FPP $\approx 49\%$).  
In addition, TIC~22069559 has an FPP of approximately 15\%, which lies near the 
critical threshold adopted in our classification (see Table~\ref{tab:pollution}).  
Given the current data quality, the origin of these transit-like signals remains 
uncertain and will require future high signal-to-noise observations for confirmation.

\subsection{Summary}

This section classifies and discusses the 18 selected candidates. Ultimately, we 
identified two highly promising candidates: TIC~48031665 and TIC~259230140. 
Although the former has an extremely shallow transit depth, it is regarded as a 
very promising candidate due to its extremely low FPP and NFPP, a clean stellar 
field, and very low contamination from nearby starlight. Although the latter 
has higher FPP and NFPP values and a greater contamination level than the former, 
its clear U-shaped transit light curve fully demonstrates its potential as an exoplanet.

Among the remaining targets, some have relatively low FPP and NFPP values, 
but their light curves show a blurred V-shaped or transitional form, indicating 
that their planetary nature remains uncertain.For some other targets, we failed to
detect reliable transit events due to weak transit signals, high contamination 
rates, or the possible misidentification of stellar activity as transit signals. 
In addition, the FPP of several sources ranges from 15\% to 50\%, with relatively 
shallow signals. Future higher-quality data are still needed for verification.

\section{Conclusion and Discussion}

In this study, we selected 18 potential exoplanet candidates orbiting A-type 
stars from the NASA Exoplanet Archive and conducted a systematic analysis of 
their contamination from nearby stars, FPP, and NFPP. The results indicate 
that the signals of some candidates may be obscured by noise 
or may represent misidentified stellar activity interpreted as transit signals 
(such as two candidates 
labeled as APC), for which no corresponding transit signals were found at 
the TOI periods reported in the database.  
Among the candidates showing possible transit-like signals, 
we identified two particularly noteworthy sources, TIC~48031665 and TIC~259230140. 
Although the transit signal of the former is extremely shallow, its very low FPP ($\approx2.3\%$) 
and clean stellar field (with an average contamination rate of only $\approx0.5\%$) 
strongly suggest that it is a genuine exoplanet. 
The latter, although having a higher FPP ($\approx7\%$) and contamination rate 
($\approx7\%$) than the former, 
exhibits a well-defined U-shaped transit profile with appreciable depth, 
characteristic of a typical exoplanet and is likewise a highly credible candidate.
Other targets show shallower or ambiguous V-shaped features. 
Although some of these exhibit low FPP values, their light curve morphologies 
resemble those of eclipsing binaries, implying that the model may still 
produce misclassifications. Therefore, FPP should only serve as an auxiliary 
indicator of the likelihood of a true planetary nature. When its value lies 
within the range of 10\%---90\%, the credibility of the result may be misleading, 
and additional diagnostics---such as radial velocity measurements and other 
complementary analyses---are required for reliable validation.

In the computation of FPP and NFPP, we adopted phase-folded light curves, 
whose reliability depends on the accuracy of the detrending process and 
the precision of the orbital period. To reduce computational cost, we used 
orbital parameters derived from MAP optimization 
rather than those from the full MCMC posterior distributions. This approach 
limits the ability to quantify the uncertainties of orbital parameters and 
thus the resulting distributions of FPP and NFPP under input errors. 
However, as shown in Table~\ref{tab:map_mcmc_params}, the MAP and MCMC 
results are generally consistent. Moreover, when FPP falls within the 
10\%---90\% range, its reliability is inherently low, as discussed above. 
Therefore, using MAP-derived parameters for rapid estimation has a 
negligible impact on the overall conclusions. This strategy facilitates 
the early identification of high-value candidates, while subsequent 
confirmation of the transit signal sources can be achieved through more 
robust MCMC analyses.

The scope of this study includes only A-type star candidates with orbital 
periods longer than one day and planetary radii smaller than $7\,\text{R}_\oplus$, 
which may introduce certain selection biases. To avoid contamination from 
eclipsing binaries, we excluded candidates with periods shorter than one 
day, which may have consequently led to the omission of potential USPs. 
Meanwhile, to eliminate possible brown dwarf scenarios, an upper radius 
limit of $7\,\text{R}_\oplus$ was applied, leading to the exclusion of larger 
planetary candidates such as Jupiter-sized planets. Future studies are 
encouraged to include candidates with shorter periods and larger radii 
to comprehensively assess the distribution characteristics of exoplanets 
around A-type stars.

\begin{acknowledgments}
    
This work is supported by the National Natural Science Foundation of China (No.~12573038), 
the Yunnan Fundamental Research Projects 
(Grant Nos.~202503AP140013,~202401AS070046 and 202501AS070055), 
the China Manned Space Program with grant no. CMS-CSST-2025-A16, 
the 2022 CAS“Light of West China”Program (Fund recipient L.W.P).
The \textit{TESS} data presented in this paper were obtained 
from the Mikulski Archive for Space 
Telescopes (MAST) at the Space Telescope Science Institute (STScI). STScI is 
operated by the Association of Universities for Research in Astronomy, Inc. 
Support to MAST for these data is provided by the NASA Office of Space Science. 
Funding for the \textit{TESS} mission is provided by the NASA Explorer Program. 
The authors are sincerely grateful to an anonymous referee for instructive advice 
and productive suggestions.

This work made use of \texttt{TESS-cont} 
(\url{https://github.com/castro-gzlz/TESS-cont}), 
which also made use of \texttt{tpfplotter} 
\citep{tess-cont_ARTICLE} and \texttt{TESS-PRF} \citep{tess-cont_software}.
\end{acknowledgments}

\appendix

\setcounter{table}{0}
\renewcommand{\thetable}{A\arabic{table}}
\setcounter{figure}{0}
\renewcommand{\thefigure}{A\arabic{figure}}

\section{Summary of Candidate Sample}
\label{tab:sample}

\startlongtable
\begin{deluxetable*}{lccccccccc}
\tablecaption{Summary of the A-type star candidate samples. \label{tab:pc_summary}}
\tablehead{
\colhead{TOI} & \colhead{TIC ID} & \colhead{R.A. (deg)} & \colhead{Decl. (deg)} &
\colhead{$T_{\mathrm{eff}}$ (K)} & \colhead{$R_\star$ ($R_\odot$)} & \colhead{$M_\star$ ($M_\odot$)} &
\colhead{$T$ (mag)} & \colhead{Period (days)} & \colhead{$R_p$ ($R_\oplus$)}
}
\startdata
7183.01 & 422159302 & 335.6231 & 58.0831 & 9088 & 2.30 & \nodata & 9.50 & 10.5739 & 4.61 \\
4550.01 & 400972123 & 167.8198 & -68.4725 & 7751 & 2.57 & 2.26 & 9.77 & 1.8869 & 4.83 \\
1981.01 & 378613125 & 155.5906 & -63.6546 & 8749 & 2.14 & 2.49 & 10.00 & 2.4763 & 4.73 \\
6915.01 & 367439938 & 78.3324 & 31.3758 & 8013 & 1.98 & 1.94 & 10.57 & 2.4916 & 5.47 \\
4386.01 & 350575997 & 246.6214 & -63.4579 & 8127 & 1.75 & 1.98 & 9.77 & 2.0134 & 5.85 \\
4510.01 & 340458804 & 115.6390 & -58.6233 & 7923 & 1.52 & 1.90 & 10.36 & 194.2433 & 4.47 \\
993.01  & 259353953 & 113.9608 & -15.4995 & 8035 & 1.67 & 1.94 & 10.06 & 1.4635 & 4.76 \\
4384.01 & 259230140 & 223.1936 & -68.6665 & 7596 & 1.61 & 1.77 & 9.66 & 14.3172 & 6.38 \\
2448.02 & 231721005 & 80.7402 & -53.7646 & 9671 & 1.78 & 2.47 & 10.00 & 19.7580 & 3.81 \\
989.01  & 207277638 & 110.5065 & 10.2386 & 7814 & 2.06 & 1.88 & 10.11 & 3.1212 & 5.74 \\
6262.01 & 180575165 & 129.1580 & -36.5232 & 9532 & 1.75 & 2.43 & 9.96 & 1.4946 & 2.79 \\
5387.01 & 166086403 & 214.4454 & 62.7584 & 7531 & 1.52 & 1.74 & 8.20 & 2.8038 & 1.27 \\
4180.01 & 126737992 & 136.9193 & -50.6402 & 7511 & 1.92 & 1.75 & 8.29 & 1.5954 & 4.32 \\
6260.01 & 60984804  & 140.5958 & -6.0606  & 7707 & 1.96 & 1.81 & 7.12 & 2.3898 & 2.33 \\
2035.01 & 53709089  & 17.7367  & 61.0921  & 9387 & 2.35 & \nodata & 9.12 & 8.5123 & 5.90 \\
4588.01 & 48031665  & 282.0004 & 48.5423  & 7800 & 1.57 & 1.85 & 9.07 & 13.1188 & 2.99 \\
4347.01 & 32498058  & 84.1936  & -24.0312 & 7607 & 1.56 & 1.78 & 8.83 & 35.3573 & 4.03 \\
4501.01 & 22069559  & 292.7237 & 30.4335  & 8279 & 2.94 & 2.04 & 8.47 & 2.0349 & 3.58 \\
\enddata
\tablecomments{
$R_\star$ ($R_\odot$) and $M_\star$ ($M_\odot$) are adopted from ExoFOP-TESS, 
while all other parameters are taken from the NASA Exoplanet Archive (on September~29,~2025). 
The missing stellar mass entries (for TIC~422159302 and TIC~53709089) correspond to targets 
without available mass estimates in ExoFOP-TESS.
}
\end{deluxetable*}

\bibliographystyle{aasjournal}
\bibliography{refs}

\end{document}